\documentclass[aps,pra,twocolumn,superscriptaddress]{revtex4-2}

\usepackage{graphicx}
\usepackage{float}
\usepackage{physics}
\usepackage[hidelinks]{hyperref}
\usepackage[capitalise,noabbrev]{cleveref}
\usepackage{amssymb}

\begin{document}

\newcommand{\goo}{g_{11}}
\newcommand{\gii}{g_{ii}}
\newcommand{\aii}{a_{ii}}
\newcommand{\aoo}{a_{11}}
\newcommand{\att}{a_{22}}
\newcommand{\aot}{a_{12}}
\newcommand{\gtt}{g_{22}}
\newcommand{\got}{g_{12}}
\newcommand{\Elhy}{E_{\mathrm{LHY}}}


\title{Quantum Droplets in Imbalanced Atomic Mixtures}

\author{T. A. Flynn}
\email{t.flynn@ncl.ac.uk}
\author{L. Parisi}
\author{T. P. Billam}
\author{N. G. Parker}
\affiliation{Joint Quantum Centre (JQC) Durham--Newcastle, School of Mathematics, Statistics and Physics, Newcastle University, Newcastle upon Tyne, NE1 7RU, United Kingdom}



\date{\today}

\begin{abstract}
Quantum droplets are a quantum analogue to classical fluid droplets in that they are self-bound and display liquid-like properties --- such as incompressibility and surface tension --- though their stability is the result of quantum fluctuations. One of the major systems for observing quantum droplets is two-component Bose gases. Two-component droplets are typically considered to be balanced, having a fixed ratio between the densities of the two component. This work goes beyond the fixed density ratio by investigating spherical droplets in imbalanced mixtures. With increasing imbalance, the droplet is able to lower its energy up to a limit, at which point the droplet becomes saturated with the atoms of the majority component and any further atoms added to this component cannot bind to the droplet. Analysing the breathing mode dynamics of imbalanced droplets indicates that the droplet can emit particles, as in balanced mixtures, but the imbalance leads to an intricate superposition of multiple simultaneously decaying collective oscillations.
\end{abstract}


\maketitle

\section{INTRODUCTION \label{sec:intro}}

Quantum gases are highly controllable systems due in part to their diluteness, yielding a highly versatile platform for exploring quantum many-body physics \cite{bloch2008}. For a system of bosons, a mean-field model for quantum gases is Gross-Pitaevskii (GP) theory \cite{gross1961,pitaevskii1961}, which captures much of the physics of these systems, whilst neglecting beyond-mean-field effects such as quantum fluctuations \cite{pit2016,pethick2008}. For a single-component homogeneous Bose gas stability is governed by the atomic interactions, with the gas becoming unstable to collapse for attractive interactions. This has been demonstrated experimentally by using a Feshbach resonance to tune the interactions from repulsive to attractive \cite{roberts2001,donley2001}. For a two-component system, the stability depends on both the inter- and intraspecies contact interactions, which can likewise be tuned via Feshbach resonances \cite{courteille1998,inouye1998,chin2010}. For repulsive intraspecies interactions --- to ensure stability of each individual component --- and attractive interspecies interactions, mean-field theory once more predicts an unstable gas. However, this attractive collapse can be stabilised at high densities by quantum fluctuations \cite{petrov2015} forming a self-bound quantum droplet; this has inspired work in so called Lee-Huang-Yang (LHY) fluids \cite{jorgensen2018,skov2020}, named for the correction describing quantum fluctuations to first-order. By tuning mean-field interactions between the components to vanish, the interactions of LHY fluids are described by quantum fluctuations alone \cite{jorgensen2018,skov2020}.

In free space quantum droplets exist in equilibrium with the vacuum and form a seemingly counter-intuitive dilute liquid-like state \cite{petrov2015,petrov2018}. This adds to the variety of properties that quantum gases can be used to investigate, as the majority of experiments are inherently within the gas phase --- a property widely exploited in time-of-flight imaging \cite{castin1996} --- whereas quantum droplets open up the field to explore the properties of quantum liquids, such as surface tension \cite{petrov2015,ancilotto2018,2ciko2021} and incompressibility \cite{baillie2017,ferioli2019}, within controllable experimental conditions.

Quantum droplets have been experimentally observed in two systems:  (i) dipolar gases of Dy \cite{kadau2016,ferrier2016,schmitt2016} and Er \cite{chomaz2016}; (ii) homonuclear $^{39}$K \cite{cabrera2018,semeghini2018,cheiney2018,ferioli2019} and heteronuclear, $^{41}$K-$^{87}$Rb and $^{23}$Na-$^{87}$Rb \cite{derrico2019,guo2021}, two-component mixtures. The underlying mechanism for droplet formation is the result of a vanishing mean-field interspecies attraction tending to cause instability, that is countered by the repulsive quantum fluctuations which become significant with increasing density \cite{petrov2015}. However, the mean-field interactions within a dipolar Bose gas include both the two-body, short-ranged interactions of a non-dipolar gas, and anisotropic, long-ranged interactions resulting from strong atomic magnetic dipoles \cite{lahaye2009}. This anisotropy results in elongated droplet profiles \cite{wachtler2016,bisset2016,pal2020,auchen2021,pal2022}, which differentiates dipolar droplets from non-dipolar, two-component droplets.

A defining property of three-dimensional (3D) two-component quantum droplets is self-evaporation \cite{petrov2015}. In certain regimes of the droplet's phase diagram, excitations will cause the droplet to shed atoms in order to relax to a system of lower energy. This occurs when the energies of these excitations exceed $-\mu$, the particle emission threshold, where $\mu$ is the chemical potential of the droplet. Furthermore, the lowest energy monopole mode --- the breathing mode --- decays across the largest proportion of the droplet's phase diagram relative to other collective modes \cite{petrov2015,fort2021}. Self-evaporation is a non-intuitive and remarkable property that is not exhibited by dipolar quantum droplets \cite{baillie2017}, or for the breathing mode of a 1D two-component droplet \cite{tylutki2020}, and has not been experimentally observed, but it is crucial in understanding the dynamics of these objects.

The formation of a two-component quantum droplet also has an interesting property: density balancing. A key result from the pioneering work of Petrov is the energetic favourability for the two component densities to maintain a fixed ratio $n_2/n_1 = \mathrm{const.}$ \cite{petrov2015} where $n_i$ is the number density of the $i$th component. By pairing this assumption with negligible spin modes \cite{petrov2015} --- i.e., assuming only in-phase density oscillations --- the mixture can be modelled via a single macroscopic wavefunction. The majority of the literature has focused on such balanced droplets, with theoretical studies of imbalanced systems limited to dipolar mixtures \cite{smith2021,bisset2021} and low-dimensions \cite{mithun2020,tengstrand2022}. This work is a systematic study of the ground states and breathing modes of 3D spherical imbalanced quantum droplets in homonuclear mixtures. Adding atoms to one component of the mixture yields a lower energy configuration than the associated balanced droplet. This forms a droplet with a density imbalance in the core, and this imbalance can be increased to a limit at which point any further atoms cannot bind to the droplet. By investigating imbalanced droplets in the free space limit, this work explores how the density profiles, chemical potentials and breathing mode dynamics of the droplet are modified by the presence of a population imbalance.

This work begins with a discussion of the underlying theory in \cref{sec:mod}. This theory is then applied in Sections \ref{sec:gs} and \ref{sec:breath} to first isolate imbalanced droplet ground states and then to propagate these states in time, subject to an initial perturbation, to analyse the droplet breathing modes. Finally the main results and potential future research avenues are discussed in \cref{sec:disc}.

\section{THE MODEL \label{sec:mod}}
A zero-temperature mixture of two weakly-interacting, dilute Bose gases can be described by the energy functional \cite{ho1996,petrov2015}
\begin{equation}
\begin{split}
	E = \int&\bigg[\frac{\hbar^2}{2m_1}|\grad{\Psi_1}|^2 + \frac{\hbar^2}{2m_2}|\grad{\Psi_2}|^2 + \frac{\goo}{2}|\Psi_1|^4 + \frac{\gtt}{2}|\Psi_2|^4 \\
&+ \got|\Psi_1|^2|\Psi_2|^2 + \frac{\Elhy}{V} \bigg]\dd^3\mathbf{r},
\end{split}
\label{eq:e_func}
\end{equation}
in which $m_i$ are the atomic masses of the $i$-th component, $g_{ii}$ and $\got$ are the effective intra- and interspecies interaction strengths respectively, and are related to the intra- and interspecies scattering lengths by $g_{ii} = 4\pi\hbar^2a_{ii}/m_i$ and $\got = 2\pi\hbar^2\aot(1/m_1 + 1/m_2)$. The first two terms of \cref{eq:e_func} are the kinetic energy contributions; the next three terms describe the two-body interactions within and between the components. The final term is the LHY correction which, to first-order, describes the effects of quantum fluctuations on the condensate \cite{lee1957}. For a Bose-Bose mixture, the LHY correction takes the form \cite{petrov2015}
\begin{equation*}
\frac{\Elhy}{V} = \frac{8}{15\pi^2}\left(\frac{m_1}{\hbar^2}\right)^{3/2}\left(\goo n_1\right)^{5/2}f\left(z,u,x\right),
\end{equation*}
where $f$ is a function, defined in \cite{ancilotto2018,minardi2019}, with arguments $z = m_1/m_2$, $u = \got^2/(\goo\gtt)$ and $x = (\gtt n_2)/(\goo n_1)$. The function $f(z,u,x)$ means that the LHY term takes a relatively simple form in homonuclear mixtures ($m_2 = m_1$) \cite{petrov2015}, but takes a far more complex form in heteronuclear mixtures ($m_2 \neq m_1$) \cite{petrov2015,ancilotto2018}. The function argument, $u$, reduces to one by assuming that the mixtures lies at the critical point of attractive stability, i.e., $\got^2 = \goo\gtt \implies u = 1$, removing the issue of complex contributions resulting from an unstable phonon mode \cite{petrov2015,qi2020,xiong2022}. It should be noted that this approximation is made only in this derivation and does not imply any parameter choice in subsequent analyses. Crucial to the formation of quantum droplets, quantum fluctuations stabilise the attractive mixture against collapse, and furthermore are ubiquitous in nature though often play a limited role in the physics of many quantum gas experiments as discussed in \cref{sec:intro}.

The two component densities can each be related to a macroscopic wavefunction or order parameter, $n_i = |\Psi_i|^2 = \Psi_i^{\ast}\Psi_i$. A variational approach, $i\hbar\pdv*{\Psi_i}{t} = \fdv*{E}{\Psi_i^{\ast}}$ \cite{pit2016}, can then be used to derive the coupled extended GP equations,
\begin{align}
\begin{split}
&i\hbar\pdv{\Psi_1}{t} = \bigg[-\frac{\hbar^2}{2m_1}\laplacian + \goo|\Psi_1|^2 + \got|\Psi_2|^2 \\
&+ \frac{8}{15\pi^2}\left(\frac{m_1}{\hbar^2}\right)^{3/2}\goo^{5/2}\left\{\frac{5}{2}f|\Psi_1|^3 - \frac{\gtt}{\goo}\pdv{f}{x}|\Psi_1||\Psi_2|^2\right\}\bigg]\Psi_1, \\
&i\hbar\pdv{\Psi_2}{t} = \bigg[-\frac{\hbar^2}{2m_2}\laplacian + \gtt|\Psi_2|^2 + \got|\Psi_1|^2 \\
&+ \frac{8}{15\pi^2}\left(\frac{m_1}{\hbar^2}\right)^{3/2}\goo^{3/2}\gtt\pdv{f}{x}|\Psi_1|^3 \bigg]\Psi_2.
\label{eq:gen_gpes}
\end{split}
\end{align}
The computational complexity of these LHY terms can be minimised by assuming a homonuclear mixture, giving the equal-mass coupled extended GP equations \cite{petrov2015}
\begin{align*}
i\hbar\pdv{\Psi_1}{t} = &\bigg[-\frac{\hbar^2}{2m}\laplacian + \frac{4\pi\hbar^2\aoo}{m}|\Psi_1|^2 + \frac{4\pi\hbar^2\aot}{m}|\Psi_2|^2 \\
&+ \frac{128\sqrt{\pi}\hbar^2\aoo}{3m}\left(\aoo|\Psi_1|^2 + \att |\Psi_2|^2\right)^{3/2}\bigg]\Psi_1, \\
i\hbar\pdv{\Psi_2}{t} = &\bigg[-\frac{\hbar^2}{2m}\laplacian + \frac{4\pi\hbar^2\att}{m}|\Psi_2|^2 + \frac{4\pi\hbar^2\aot}{m}|\Psi_1|^2 \\ 
&+ \frac{128\sqrt{\pi}\hbar^2\att}{3m}\left(\aoo|\Psi_1|^2 + \att |\Psi_2|^2\right)^{3/2}\bigg]\Psi_2.
\end{align*}
Finally, the dimensional scalings $\mathbf{r} = \xi\mathbf{\tilde{r}}$, $t=\tau\tilde{t}$ and $\Psi_i = \rho_i^{1/2}\tilde{\Psi}_i$ result in the dimensionless, equal-mass coupled extended GP equations,
\begin{align}
\begin{split}
i\pdv{\Psi_1}{t} = &\bigg[-\frac{1}{2}\laplacian + |\Psi_1|^2 + \eta|\Psi_2|^2 \\
&+ \alpha\left(|\Psi_1|^2 + \beta|\Psi_2|^2\right)^{3/2}\bigg]\Psi_1, \\
i\pdv{\Psi_2}{t} = &\bigg[-\frac{1}{2}\laplacian + \beta|\Psi_2|^2 + \eta\beta|\Psi_1|^2 \\
&+ \alpha\beta^2\left(|\Psi_1|^2 + \beta|\Psi_2|^2\right)^{3/2}\bigg]\Psi_2.
\label{eq:dim_gpes}
\end{split}
\end{align}
in which all tildes are hereafter neglected and the dimensionless parameters are 
\begin{equation*}
\eta = \frac{\aot}{\sqrt{\aoo\att}},
\end{equation*}
\begin{equation*}
\alpha = \frac{32}{3}\left[\frac{2}{3\pi}\frac{|\delta a|\aoo^{5/2}n_1^{(0)}}{\sqrt{\aoo} + \sqrt{\att}}\right]^{1/2}, \, \beta = \left(\frac{\att}{\aoo}\right)^{1/2},
\end{equation*}
with dimensional parameters
\begin{align*}
	\xi = \sqrt{\frac{3}{8\pi}\frac{(\sqrt{\aoo} + \sqrt{\att})}{|\delta a|\sqrt{\aoo} n_1^{(0)}}}, & & \tau = \frac{3m}{8\pi\hbar}\frac{(\sqrt{\aoo} + \sqrt{\att})}{|\delta a| \sqrt{\aoo} n_1^{(0)}}, \\ \rho_1 = \frac{2}{3}\frac{|\delta a|n_1^{(0)}}{\sqrt{\aoo}(\sqrt{\aoo} + \sqrt{\att})}, & & \rho_2 = \frac{2}{3}\frac{|\delta a|n_1^{(0)}}{\sqrt{\att}(\sqrt{\aoo} + \sqrt{\att})}, \\
\end{align*}
where $\delta a = \aot + \sqrt{\aoo\att}$, and $n_1^{(0)}$ is the equilibrium density of component-1 for the balanced mixture. The expression of the equilibrium density is calculated in a homogeneous infinite system under the criterion of a vanishing pressure --- i.e., the droplet in equilibrium with the vacuum --- and takes the form \cite{petrov2015} 
\begin{equation*}
n_1^{(0)} = \frac{25\pi}{1024} \frac{(\aot + \sqrt{\aoo\att})^2}{\aoo^{3/2}\att(\sqrt{\aoo} + \sqrt{\att})^5}.
\end{equation*}
The density scalings $\rho_i$ correspond to rescaled normalisation constants, $\tilde{N_i} = N_i/(\rho_i\xi^3)$, in which $N_i$ is the normalisation constant of the $i$-th component wavefunction. In subsequent sections the dimensionless forms of $N_i$, $\Psi_i$, $\mathbf{r}$ and $t$ are used, with tildes omitted for clarity.

By assuming a constant density ratio, $n_2/n_1 = \sqrt{\goo/\gtt}$, the two component wavefunctions can be expressed in terms of a single wavefunction, $\Psi_i = \sqrt{n_i}\phi$, neglecting any out-of-phase motion between the components \cite{petrov2015,bienaime2016,capellaro2018}. Equations (\ref{eq:gen_gpes}) reduce to a single equation,
\begin{equation*}
i\pdv{\phi}{t} = \left[-\frac{1}{2}\laplacian - 3|\phi|^2 + \frac{5}{2}|\phi|^3\right]\phi,
\end{equation*}
with the system described by a single parameter, an effective atom number, $\tilde{N}$, given by \cite{petrov2015}
\begin{equation}
\tilde{N} = \left(\frac{\sqrt{\gtt}}{n_{1}^{(0)}(\sqrt{\goo} + \sqrt{\gtt})}\right)\frac{N}{\xi^3},
\label{eq:eff_num}
\end{equation}
in which $N$ here is the total atom number $N = N_1 + N_2$. Within this work, balanced and imbalanced droplets are both modelled by Equations (\ref{eq:dim_gpes}), though it should be noted that, for a balanced droplet, the dimensionless parameters $(N_1, N_2,\alpha,\beta,\eta)$ can be recast to $\tilde{N}$. In the density-locked model, a given set of interaction strengths, $g_{ii}$ and $g_{12}$, correspond to a fixed population number ratio, $N_2/N_1 = \sqrt{\goo/\gtt}$, imposing that the atoms can only reside bound to the droplet. By breaking the assumption of density-locking, it is possible to imbalance population numbers such that $N_2/N_1 \neq \sqrt{\goo/\gtt}$. The next section explores the effect of this imbalance on the density structure and energy of 3D spherical droplet ground states.

\section{GROUND STATES \label{sec:gs}}

This work considers spherically-symmetric droplets. Density is hence assumed to be a function of radius only, reducing the computational problem to an effective 1D system with the kinetic term becoming $\laplacian\Psi_i \rightarrow [\pdv*[2]{(r\Psi_i)}{r}]/r$. Further assumptions within this work include: a homonuclear mixture and balanced intraspecies scattering lengths ($\aoo = \att \implies \beta = 1$), to simplify the problem. Thus, the only differences between the components can arise from an imposed atom number imbalance of $N_1 = N_2 + \delta N_1$.

\subsection{Flat-top Droplet Limit}

\begin{figure}
  \centering
  \includegraphics[width=0.5\textwidth]{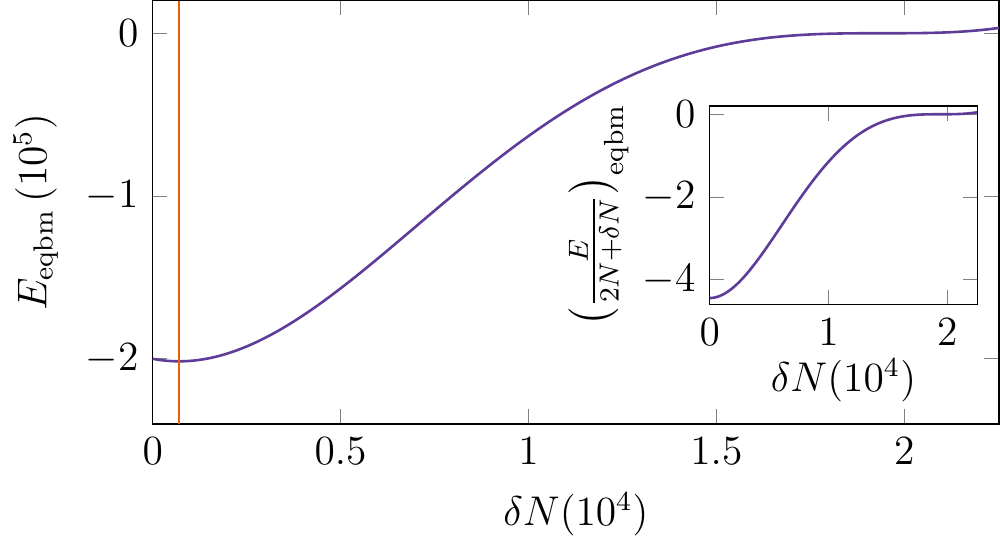}
  \caption{Energy of a flat-top density ansatz as a function of imbalance, $\delta N$, with parameters $N_2\approx22524$, $\alpha\approx0.0152$, $\beta=1.0$ and $\eta=-1.2$. The equilibrium energy, $E_{\mathrm{eqbm}}$, is shown in the main plot with the orange, vertical line indicating the location of the minimum at $\delta N \approx 703$. The inset shows the equilibrium energy per particle, $[E/(2N+\delta N)]_{\mathrm{eqbm}}$, with the only minimum appearing at the origin.}
  \label{fig:var_fig}
\end{figure}

A simplified ansatz for the density of large droplets, as used in Ref.~\cite{petrov2015}, is a flat-top, step function in which kinetic energy contributions are neglected. The two-components of the imbalanced droplet are modelled as majority and minority components with $N+\delta N$ and $N$ atoms, respectively. Substituting this ansatz into the two-component energy functional gives the equilibrium energy of the flat-top droplet, $E_{\mathrm{eqbm}}$, as a function of the imbalance, $\delta N$ (see the \cref{sec:var_deriv}for further details, including a schematic of the ansatz). The results of this energy calculation are given in \cref{fig:var_fig} with the total equilibrium energy given in the main plot, and the inset showing the equilibrium energy per particle.

Firstly, the main plot of \cref{fig:var_fig} shows that there is a minimum at
$\delta N \neq 0$ in the total energy, indicating the droplet can lower its
energy by absorbing particles into one component. However, the equilibrium
energy per particle given in the inset of \cref{fig:var_fig} exhibits a minimum
at $\delta N=0$ demonstrating that whilst the total energy of the droplet can
be lowered by single-component absorption, there is a corresponding increase in
the energy per particle resulting in a less stably bound droplet. Secondly, the
\cref{sec:var_deriv}shows that the minimum in the equilibrium energy occurs at
$\delta N / N = 2 (\eta - 2) + [(\eta - 1) (4 \eta - 14)]^{1/2}$. That is, the
energy of the droplet decreases with imbalance up to this limit. The dependence
of this quantity on the scattering lengths is consistent with the prediction of
Ref.~\cite{petrov2015} at linear order \footnote{Ref.~\cite{petrov2015} finds
  $\delta N / N \sim |\delta g| / g = -(1+\eta)/2$ for $|\delta g| / g \ll 1$;
expanding our result in the same limit yields $\delta N / N \approx (1/3)
|\delta g| / g$.}. For the parameters of \cref{fig:var_fig} $\delta
n\approx0.458$, as highlighted by the orange vertical line.

 At this limit the droplet becomes saturated with the
majority component. Hence if further majority component atoms are available,
they cannot be absorbed into the droplet and will form an unbound gas around
the droplet (as described in Ref.~\cite{petrov2015}). There are thus three
droplet states: (1) balanced droplets; (2) bound, imbalanced droplets; and (3)
saturated, imbalanced droplets (which can be immersed in a halo of unbound
atoms). 

The decrease in energy of an imbalanced droplet is consistent with the predictions of Ref.~\cite{petrov2015} and stems from the basic argument that droplets always seek to lower their energy by absorbing atoms. For example, in the density-locked model, atoms can only be added to both components to ensure that $N_2/N_1 = \sqrt{\gtt/\goo}$ is preserved, and by adding atoms to both components, a lower energy state is recovered. One way to conceptualise this is due to droplets having negative chemical potentials, meaning that it is always energetically favourable to absorb more particles into both components. However, the subtlety in the energy calculation given in the \cref{sec:var_deriv}is that the energy of a droplet can also be lowered by absorbing particles into one component if there is an excess of this component available, up to the saturation limit discussed above.

By using the simple flat-top density ansatz, key insights are gained into effects of a population imbalance on the density structure and energy of a two-component droplet. However, for smaller droplet sizes kinetic energy cannot be neglected; thus, to extend this analysis to general, spherical droplets, the coupled GP equations are solved numerically.

\subsection{Numerical Solutions}

\begin{figure*}
\begin{center}
  \begin{tabular}{ccc}
    \includegraphics[width = 0.33\textwidth]{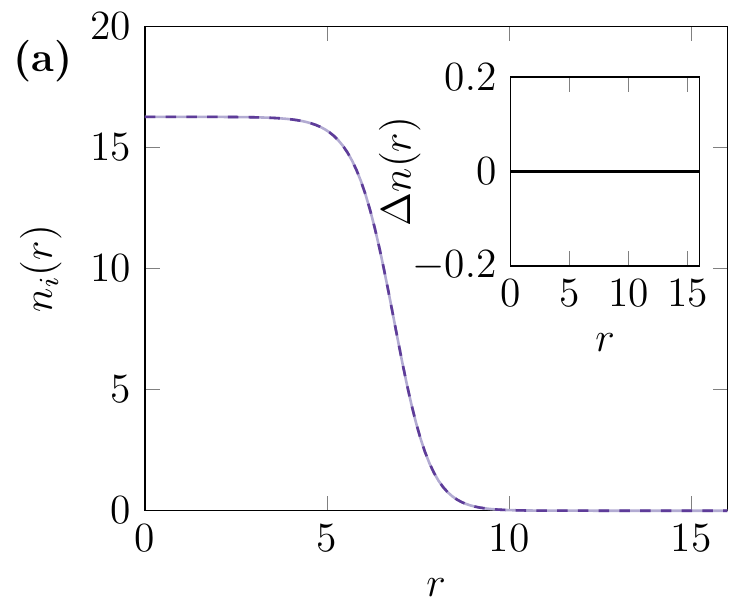} &
    \includegraphics[width = 0.33\textwidth]{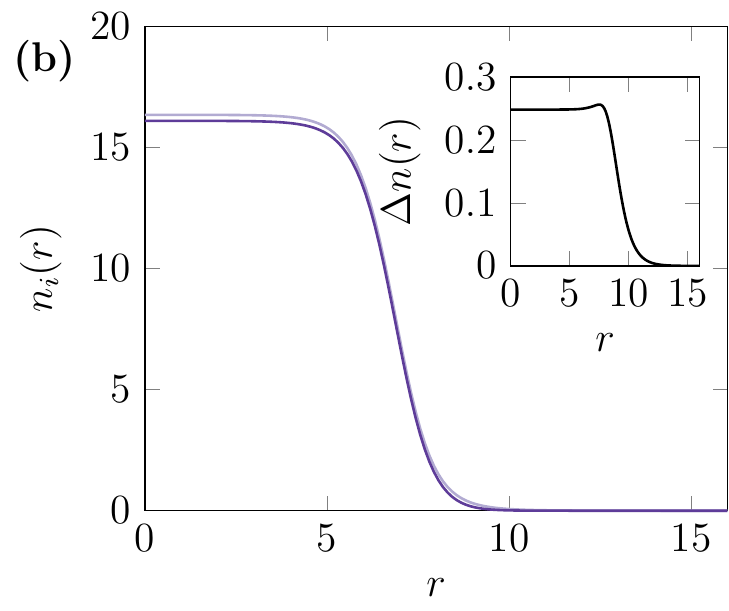} &
    \includegraphics[width = 0.33\textwidth]{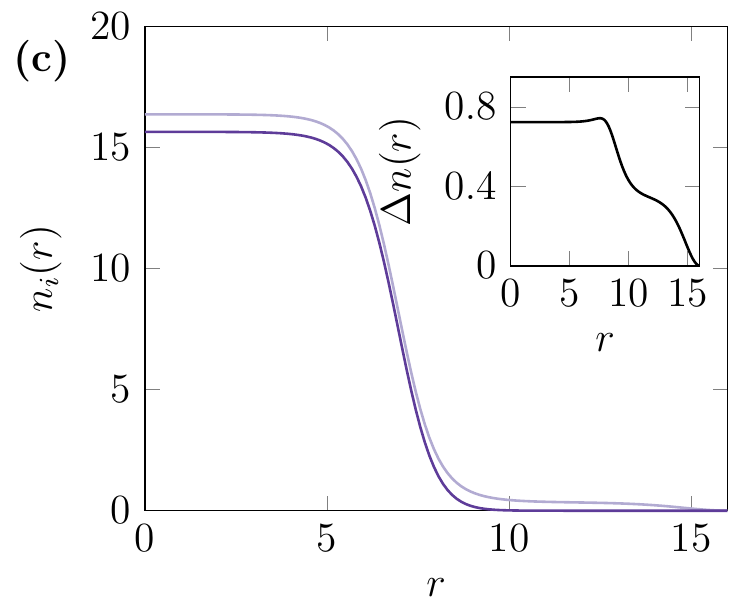}
  \end{tabular} \\
  \begin{tabular}{cc}
    \includegraphics[width = 0.5\textwidth]{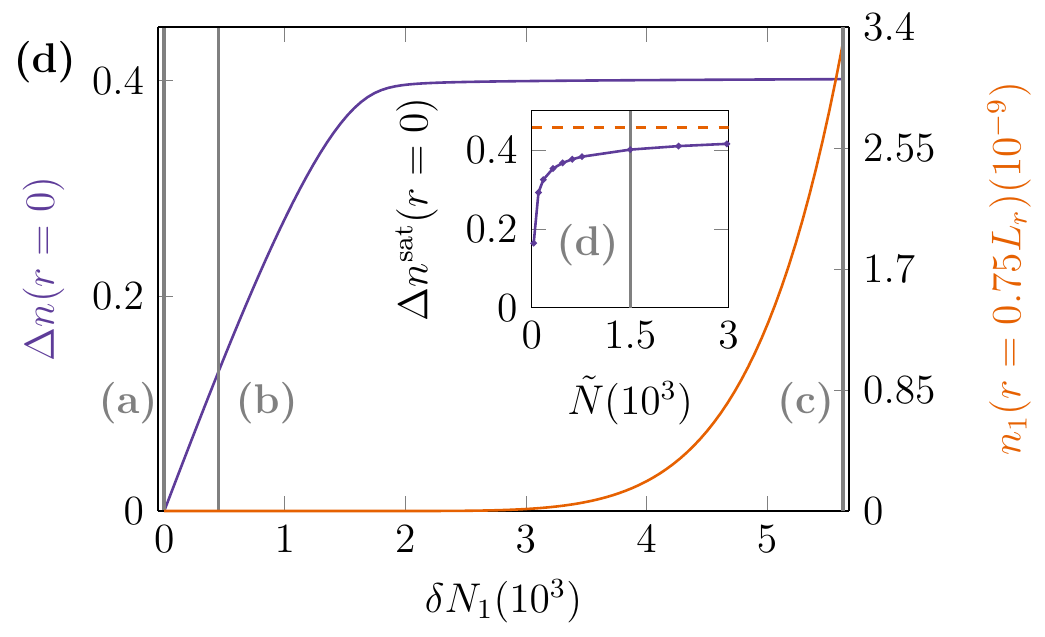} &
    \includegraphics[width = 0.5\textwidth]{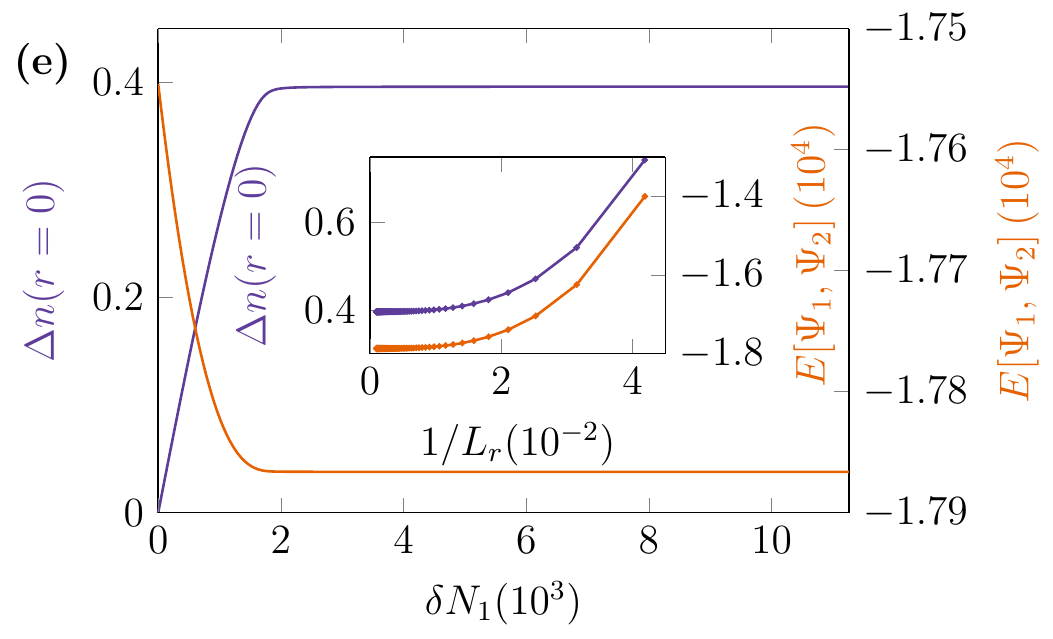}
  \end{tabular}
\end{center}
\caption{Balanced and imbalanced ground state droplets, with fixed parameters: $N_2 \approx 22524$, $\alpha \approx 0.0152$, $\beta = 1.0$ and $\eta = -1.2$. (a), (b) and (c) Ground state density profiles of: a balanced droplet ($\delta N_1 = 0$); a bound, imbalanced droplet ($\delta N_1 \approx 450$); and a saturated, imbalanced droplet with an unbound halo of majority component atoms ($\delta N_1\approx 5631$), respectively. The inset shows the difference in component densities, $\Delta n(r)$. (d) The difference in central densities as a function of imbalanced majority component atoms, $\delta N_1$, with box size of $L_r=128$. At $\delta N_1=0$ the system is balanced, with the leftmost vertical grey line corresponding to (a). Increasing $\delta N_1$ leads to an approximately linear increase in $\Delta n(r=0)$, there are however no atoms outside of the droplet as $n_1(r=0.75L_r)$, a measure of density outside of the droplet, is zero. This regime corresponds to a bound, imbalanced droplet with the central vertical grey line corresponding to (b). The linear increase in $\Delta n(r=0)$ eventually plateaus to a regime in which the central densities in the droplet cannot imbalanced further and the imbalanced droplet saturates with the density outside of the droplet increasing, resulting in a non-zero density of unbound atoms. The inset shows a measure of the difference in central densities in the limit of a saturated imbalanced droplet, $\Delta n^{\mathrm{sat}}(r=0)$, as a function of droplet size given by the effective atom number $\tilde{N}$ in \cref{eq:eff_num}, using the parameters for the $\delta N_1 = 0$ case. The orange, dashed, horizontal line corresponds to $\delta n\approx0.458$, with the droplet size used in the main plot highlighted by the grey, vertical line. (e) Neumann boundary conditions are applied at both boundaries of the computational box, with $L_r=1024$. The main plot shows the difference in central densities, recreating the results from (d), showing that the behaviour is not a function of the boundary conditions. It likewise demonstrates that the saturated droplet reaches a lower energy, $E[\Psi_1, \Psi_2]$, state than the balanced droplet. The inset shows an example imbalance of $\delta N_1 \approx 22524$, with varying box size $L_r$. In the limit of large a computational box any background gas will be effectively zero density, thus converging to a saturated droplet in free space.}
 \label{fig:gs_profs}
 \end{figure*}

To find ground state solutions, the coupled GP equations are propagated numerically in imaginary time until the energy of the mixture is deemed adequately converged. The numerical scheme is a $4^{\mathrm{th}}$-order Runge-Kutta method, with centred finite-difference methods for the spatial derivatives. Neumann boundary conditions ($\pdv*{\Psi_i}{r} = 0$) are applied at the centre of the computational box as the density in the droplet core is approximately constant. The boundary at $r=L_r$, where $L_r$ is the radial computational box size, has either Neumann or Dirichlet [$\Psi_i(r=L_r) = 0$] boundary conditions, which are discussed further below with reference to \cref{fig:gs_profs}.

By enforcing the Dirichlet boundary condition at $r=L_r$, \cref{fig:gs_profs}(a), (b) and (c) show the three different ground state solutions (discussed above) by varying $\delta N_1$, i.e., by varying the imbalance of the mixture. \cref{fig:gs_profs}(a) presents an example of a balanced droplet ($N_1=N_2 \implies \delta N_1 = 0$) with an inset of the density difference $\Delta n(r) = n_1(r)-n_2(r)$, showing that the two component densities are identical. By increasing the population imbalance to $\delta N_1 \approx 450$, as shown in \cref{fig:gs_profs}(b), the two component densities start to split within the droplet core, with both an increase and decrease in the majority and minority components' central densities, respectively. Hence, the two components are no longer identical but notably the density differences in the inset shows that this imbalanced droplet has no unbound atoms around the droplet. This implies that this state is a bound, imbalanced droplet. Driving the imbalance higher to $\delta N_1 \approx 5631$, \cref{fig:gs_profs}(c) demonstrates a more pronounced density splitting within the droplet core, but crucially also exhibits a non-zero gas density outside of the droplet [highlighted in the inset of (c)]. This ground state hence corresponds to a saturated, imbalanced droplet at the centre of the box with an unbound gas cloud outside of the droplet core.

For a set droplet size --- i.e., fixed $\alpha$, $\eta$ and $N_2$ --- the only free parameter is $\delta N_1$ and thus \cref{fig:gs_profs}(d) shows two measures of the droplet ground states for varying $\delta N_1$. The two measures are the central density difference, $\Delta n(r=0) = n_1(r=0) - n_2(r=0)$, and the majority component density, $n_1$, at a fixed radius outside of the droplet. The chosen fixed radius is $r = 0.75L_r$, which is chosen to be relatively far from the droplet surface but not too close to $r=L_r$, to avoid the density drop resulting from the Dirichlet boundary condition. These measures are used to illustrate the transition between bound, imbalanced droplets, and saturated, imbalanced droplets. \cref{fig:gs_profs}(d) shows that for increasing $\delta N_1$ from zero, there is an immediate density splitting within the droplet corresponding to the approximately linear increase in $\Delta n(r=0)$. It should be noted that in this regime, $n_1(r=0.75L_r)$ stays fixed at zero, meaning that this relatively small $\delta N_1$ regime corresponds to bound, imbalanced droplets. However, increasing $\delta N_1$ further eventually leads to the formation of a shoulder in $\Delta n(r=0)$ which is the saturation of the imbalanced droplet, i.e., the droplet is reaching a limit of how many majority component atoms can be absorbed. Beyond this shoulder $\Delta n(r=0)$ then approaches a constant value [labelled here as $\Delta n^{\mathrm{sat}}(r=0)$]. Once in the saturated droplet regime, there is a corresponding increase in $n_1(r=0.75L_r)$, indicating that the excess atoms reside outside of the droplet. The phase diagram in \cref{fig:gs_profs}(d) corresponds to the three ground state densities in (a), (b) and (c) given by the vertical grey lines. Furthermore, the inset of \cref{fig:gs_profs}(d) illustrates how the central density splitting of the saturated droplet varies with droplet size, given by $\tilde{N}$, the effective atom number of the balanced droplet in \cref{eq:eff_num} (from Ref.~\cite{petrov2015}). To vary droplet size, the fixed parameters are $\alpha$ and $\eta$, thus $N_2$ (and hence $N_1$) is varied for the different droplet sizes, with $\tilde{N}$ calculated for the balanced droplet $(\delta N_1 = 0)$ parameters. The central density difference of the saturated droplet increases with droplet size and asymptotically approaches a constant value of $\delta n\approx 0.458$ [given by the orange, dashed, horizontal line in the inset of (d)] in the limit of an large, flat-top droplet as discussed above.

 \cref{fig:gs_profs}(c) and (d) show that in the limit of the saturated droplet, there exists the unbound gas. This work seeks to explore imbalanced droplets, in the absence of this unbound gas, to probe the physics of imbalancing the droplet in free space. Thus, rather than imposing a Dirichlet boundary condition at $r=L_r$, instead the same Neumann boundary condition used at $r=0$, is used at $r=L_r$, also. This gives a numerical approximation to a free space system. \cref{fig:gs_profs}(e) uses equivalent parameters as in (d), with a slightly increased $\delta N_1$ range, and exhibits the same central density difference behaviour as with the Dirichlet boundary conditions at $r=L_r$. From this point on, Neumann boundary conditions are applied at both boundaries. 

 \cref{fig:gs_profs}(e) also shows the energy of the mixture which uses the dimensionless form of \cref{eq:e_func}, by defining $\tilde{E} = E/\epsilon$ where $\epsilon = [4/(3\pi^2\alpha)][\hbar^2/(m\xi^2)]$, in which tildes are subsequently neglected. The energy decreases with imbalance --- as predicted by the analytic calculated presented in the \cref{sec:var_deriv}--- until again reaching the saturation limit. Note again that a saturated, imbalanced droplet cannot absorb further majority component atoms and thus the energy of the droplet does not vary with $\delta N_1$. However, this is an effect of the large computational boxes used, in that the density of the unbound atoms is negligibly small such that once in this saturation limit the system is essentially identical with changing $\delta N_1$. In smaller computational boxes --- or equivalently, exploring much larger imbalances --- the density of the background gas becomes significant [e.g. \cref{fig:gs_profs}(c)], which is explored in the inset of (e). By fixing $\delta N_1 \approx 22524$, i.e., an imbalanced droplet in the saturation limit, the size of the computational box is varied. In small boxes, the density of the unbound gas is non-zero and thus has a significant, positive energy contribution. The existence of the background gas likewise modifies the internal structure of the droplet as can be seen by the increased value of $\Delta n(r=0)$. In the limit of large $L_r$ (or equivalently small $1/L_r$) the energy contribution of the background approaches zero as the density of the gas approaches zero. Thus in the limit of a large box, all of the ground states converge to the saturated droplet in free space, as can be seen from the convergence of both the energy and $\Delta n(r=0)$ in the inset of \cref{fig:gs_profs}(e).

\begin{figure}
\includegraphics[width=0.5\textwidth]{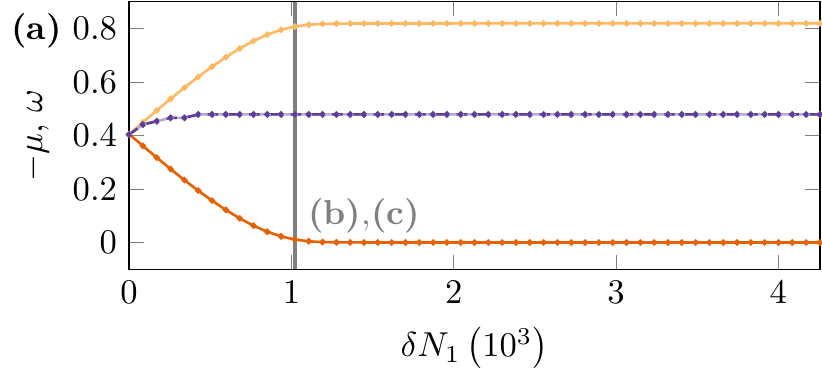}
\includegraphics[width = 0.5\textwidth]{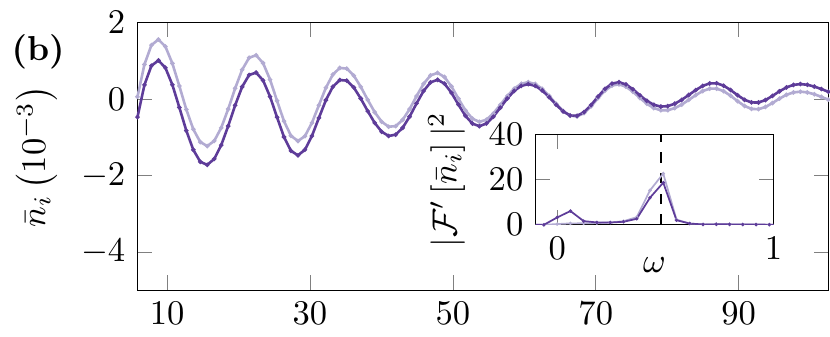} \\
\includegraphics[width=0.5\textwidth]{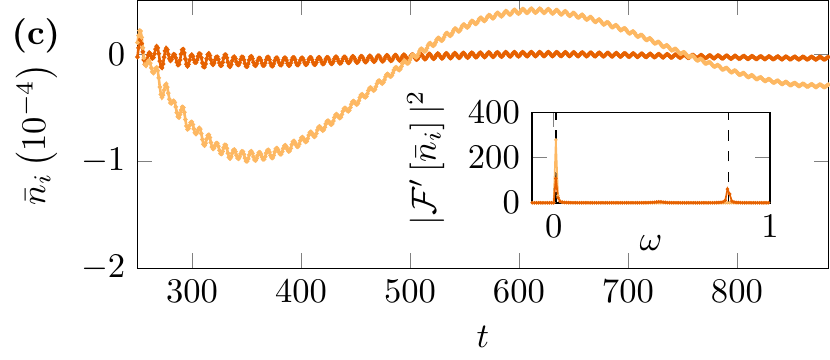} \\
\caption{Chemical potentials and breathing modes as a function of majority component imbalance, $\delta N_1$, in the self-evaporative regime (equivalent to a balanced droplet, $N_1=N_2$, with $\tilde{N} \approx 649$). (a) The chemical potentials (light orange --- minority component, dark orange --- majority component) and early-time breathing mode frequencies (light and dark purple dashed lines) as a function of imbalance. These results correspond to the range $0 \leq \delta N_1 \lesssim 4257$, $N_2 \approx 17027$, $\alpha\approx0.00657$ and $\eta\approx-1.11$. At $\delta N_1 = 0$, the chemical potentials are equal but diverge as the imbalance is increased eventually reaching a saturation point where the chemical potential of the majority component is approximately zero. The minority component chemical potential likewise saturates, to an increased value. (b) A highlighted simulation of $\delta N_1\approx 1447$ [corresponding to the bold vertical line in (a)] at early times, with an inset of the associated power spectrum. This indicates that at early times, there is a single rapidly decaying mode. (c) The same $\delta N_1 \approx 1447$ simulation at late times in which there is there is a superposition of two modes corresponding to the two chemical potentials, given by the vertical dashed lines in the inset power spectrum of (c).}
 \label{fig:SE}
 \end{figure}

Further evidence of the energy saturation is given by the two component chemical potentials in \cref{fig:SE}(a). For increasing $\delta N_1$ the majority component chemical potential (the dark orange curve) converges towards zero, implying there is no energetic favourability for the droplet to absorb more majority component atoms. Additionally, if more atoms of the minority component were made available to the droplet, then these atoms will likewise be absorbed into the droplet, reducing the energy of the droplet even further. 

In summary, droplets will always seek to absorb atoms to reduce the system energy. If there is an imbalance of atoms from $N_2/N_1 = \sqrt{\gtt/\goo}$ then the droplet will absorb more of the majority component but this reaches a limit in which the droplet is saturated with the majority component atoms. It should be noted that in the case of a non-zero density unbound gas, this limiting behaviour will be modified by the positive energy contribution of the majority component density tails, similar to submerged droplets observed in 1D imbalanced droplets \cite{mithun2020,tengstrand2022}. These results have been presented both from an analytic energy calculation, and from numerical simulations, the latter of which is to be used next to explore the dynamical stability of these imbalanced droplets across their parameter space.

 \section{BREATHING MODE \label{sec:breath}}

Self-evaporation is a key property of 3D two-component quantum droplets \cite{petrov2015}. Recalling that the density-locked droplet can be described by a single parameter, $\tilde{N}$, there are three main regimes of interest for the droplet collective modes: 1) all modes exceed the particle emission threshold and hence are evaporated ($\tilde{N}\lesssim94.2$); 2) the monopole mode evaporates but other non-zero angular momentum modes are stable ($94.2\lesssim\tilde{N}\lesssim933.7$); 3) the monopole is stabilised ($\tilde{N}\gtrsim933.7$) \cite{petrov2015,fort2021}. This first phase is the principal argument behind the self-evaporative property of a two-component droplet, since a finite-temperature droplet is inherently excited by contributions from, for example, the non-condensate components. Thus, these excitations are evaporated and the droplet relaxes to a lower energy configuration, akin to self-cooling \cite{petrov2015}. 

 In this work, the system is restricted to spherically symmetric droplets, meaning that the only observable mode is the breathing mode. The restriction of spherical symmetry is applied to reduce computational cost. This is due to particle shedding of self-evaporative droplets resulting in reflections from the computational box boundaries which become a substantial issue for long-time dynamics. To avoid this issue, very large box sizes --- $L_r=8192$, approximately $500$ times the droplet sizes considered here --- are used to observe the dynamics of the droplets without interference from reflected particles. If simulating a general 3D droplet, these box sizes would quickly become infeasible. Note that the use of large computational boxes is crucial to the focus of this paper, namely imbalanced droplets in free space, i.e., without the external effects of unbound atoms. By excluding any non-zero angular momentum modes, the spherical symmetry yields two regimes: a decaying and a stable breathing mode, thus this work analyses the stability of breathing modes in the presence of an imbalance. 

 The breathing mode frequencies of the balanced, density-locked droplet are a function of the single parameter, $\tilde{N}$ \cite{petrov2015}. However, for an imbalanced system, a further degree of freedom is introduced, the size of the imbalance. As shown in the previous section, the central density differences, chemical potentials and energies, as functions of imbalance, reach the saturation limit and any further imbalance has no significant effect on the internal structure of the droplet. In order to observe these collective modes, the ground state solutions found via imaginary time propagation, shown in \cref{sec:gs}, are then evolved in real time via Equations (\ref{eq:dim_gpes}). To trigger a breathing mode in the droplet, a harmonic phase is imprinted on the ground state wavefunction, i.e., $e^{i\epsilon r^2}$ where $\epsilon$ is small (here $\epsilon = 10^{-5}$) \cite{pit2016,stringari1996}. This phase is always imprinted onto the minority component, though breathing modes are an in phase oscillation so the subsequent dynamics are not dependent on this choice of component.

\subsection{Self-Evaporative Regime}

 The first modes to consider are within the self-evaporative regime ($\tilde{N} \lesssim 933.7$). Breathing modes of spherical, balanced droplets have been quite extensively investigated \cite{petrov2015,ferioli2020,fort2021}. Dynamically, within this regime, the droplet begins to oscillate at a frequency exceeding the particle emission threshold, $-\mu$, and rapidly decays with the decay rate asymptotically tending to zero and the oscillation frequency tending to the particle emission threshold \cite{ferioli2020}. The initial rapid decay is due to a high dissipation of energy through particle emission, though in the long-time limit this corresponds to limited particle emissions at the energy of the chemical potential.  This asymptotically decaying behaviour is likewise recovered in the imbalanced case though with three modes instead of one due to the imbalance yielding two chemical potentials. \cref{fig:SE}(a) describes the three super-imposed modes: the early-time, rapidly decaying mode --- see purple dashed lines in \cref{fig:SE}(a) denoting the extracted equal frequencies of the two components --- that varies marginally with imbalance, which can be considered as the intrinsic droplet breathing mode; at late times there are two further modes replacing the early-time mode. The late-time modes arise from the splitting of the chemical potentials which diverge with increasing imbalance before again reaching the saturated droplet [see orange lines in \cref{fig:SE}(a)]. 

To visualise the single early mode and two late modes, \cref{fig:SE}(b) and (c) represent a measure of the droplet central density, $\bar{n}_i(t) = n_i(r=0,t) - \langle n_i(r=0)\rangle_t$, with insets showing the associated power spectra $|\mathcal{F}^{\prime}[\bar{n}_i]|^2$ in which $\mathcal{F}^{\prime}[\cdot]$ denotes the power spectrum rescaled by the mean, and all negative frequencies set to zero purely for data visualisation. The early-time mode is a high-amplitude, rapidly decaying mode that is given in \cref{fig:SE}(b), with an inset of the associated power spectrum highlighting the frequency (vertical dashed line) corresponding to the purple dashed lines in \cref{fig:SE}(a). As this mode decays, it is then replaced by the two late-time modes corresponding to the split chemical potentials. The late-time dynamics is thus a superposition of a higher and lower frequency mode, as can be seen in \cref{fig:SE}(c), once more with the associated power spectra showing peaks at the two chemical potentials (given by the vertical dashed lines). It should be noted that evidence of these late-time modes can even be seen at early times, such as the shorter peak in the inset of \cref{fig:SE}(b), which roughly corresponds to the chemical potential of the majority component. 

In summary, at early times the droplet oscillates with an unstable, high-amplitude mode that decays rapidly due to energy dissipation from particle emission. In the long-time limit however, the particle emission is considerably reduced, with particles only emitted at energies of the two chemical potentials. Hence, these late-time modes decay at much slower rates than the initial mode. This is analogous to the density-locked mixture \cite{ferioli2020}, in which the mode frequency asymptotically converges to the particle emission threshold, with a vanishing decay rate. Physically this vanishing decay rate is the result of late-time emitted particles of energy $\approx -\mu$, which thus have a negligible effect on the kinetic energy of the droplet. Equivalently, for an imbalanced droplet, there are now two chemical potentials which each have associated particle emission and hence associated residual long-lived breathing modes \cite{ferioli2020}.

\subsection{Non-Self-Evaporative Regime}

 The second set of breathing modes are within the non-self-evaporative regime ($\tilde{N} \gtrsim 933.7$). For a balanced droplet in this regime the breathing mode frequency is lower than the particle emission threshold resulting in a stable oscillation \cite{petrov2015,ferioli2020}. In the self-evaporative regime, the dynamics of the imbalanced droplet is highly reminiscent of the balanced case but with further modes corresponding to the split chemical potentials, whereas the non-self-evaporative region of the droplet phase has some greater subtleties. 

\begin{figure}
\includegraphics[width = 0.5\textwidth]{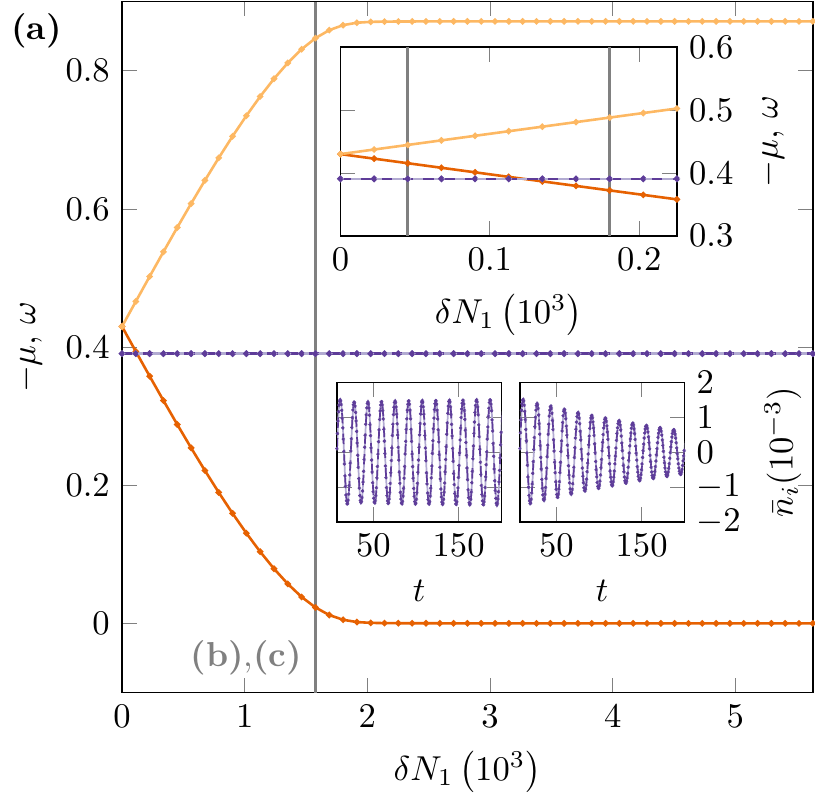}\\
\includegraphics[width = 0.5\textwidth]{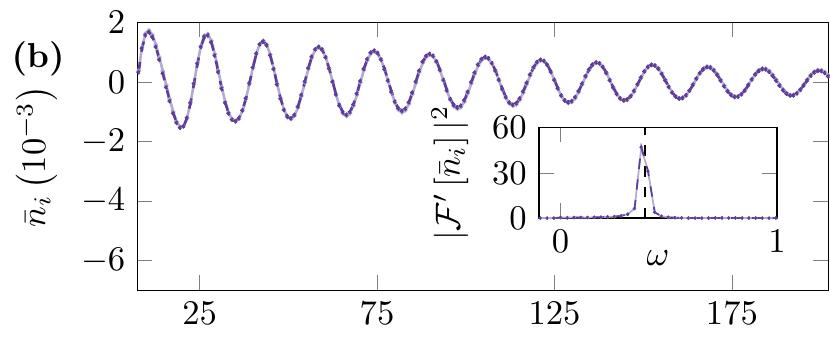}
\includegraphics[width = 0.5\textwidth]{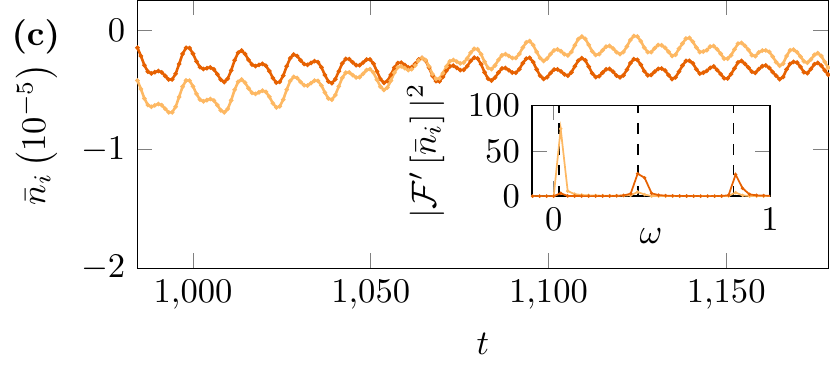} 
\caption{Chemical potentials and breathing modes as a function of majority component imbalance, $\delta N_1$, in the non-self-evaporative regime (equivalent to a balanced droplet, $N_1=N_2$, with $\tilde{N} \approx 1502$), using the same parameters as in \cref{fig:gs_profs} with $0 \leq \delta N_1 \lesssim 5631$. (a) The chemical potentials (light orange --- minority component, dark orange --- majority component) and early-time breathing mode frequencies (light and dark purple dashed lines) as a function of the imbalance. The top inset shows the majority component chemical potential crossing the breathing mode frequencies. This is shown in the lower insets in which the mode is either stable (left) or is decayed (right). (b) A highlighted simulation with $\delta N_1 \approx 1014$ [corresponding to the bold vertical line in (a)] at early times, with the associated power spectrum (inset). The early time dynamics are dominated by a single damped breathing mode. (c) The same $\delta N_1 \approx 1014$ simulation as in (b), at late times showing a superposition of three modes, with the lowest and highest energy modes corresponding to the majority and minority component chemical potentials, respectively. The third mode [the central peak in the inset of (c)] is the early-time mode still decaying.}
\label{fig:NSE}
\end{figure}

\cref{fig:NSE}(a) shows that, as in the self-evaporative case, the chemical potentials diverge with increasing imbalance until reaching the saturated, imbalanced droplet. The dynamics are again dominated by a high-amplitude mode in both components --- given by the central purple dashed lines of \cref{fig:NSE}(a) --- that is relatively constant with the changing imbalance. This mode can again be considered the intrinsic droplet breathing mode, as described in \cite{hui2020}. In the balanced case, the stability of the breathing mode for droplets of this size is due to the particle emission threshold exceeding the mode frequency. With increasing imbalance however, as the chemical potentials split, the majority component chemical potential [i.e., the lower chemical potential branch in \cref{fig:NSE}(a)] eventually crosses over the frequency of the stable high-amplitude mode, at which point this mode will begin to decay. This mode crossing implies that the imbalanced non-self-evaporative regime can instead be split into two regions: 1) a stable breathing mode; 2) a decaying breathing mode. This behaviour is highlighted in the upper inset of \cref{fig:NSE}(a), focusing on smaller imbalances. Decay of the high-amplitude, intrinsic mode occurs when the frequency exceeds the negative of the majority component chemical potential, validating the idea that stability of the mode is entirely dependent on the mode frequency exceeding the droplet's particle emission threshold. The lower left inset of \cref{fig:NSE}(a) shows that the mode is stable, for an imbalance of $\delta N_1 \approx 45$, as the frequency lies beneath both chemical potentials. However, the lower right inset shows that this mode becomes unstable, for an imbalance of $\delta N_1 \approx 180$, and decays as the frequency exceeds one of the chemical potential branches. The critical imbalance between stability and instability of this mode is $\delta N_1 \approx 124$, for this specific droplet. Thus for a sufficiently small imbalance, it is possible to have an imbalanced droplet with a stable breathing mode.

Figure \ref{fig:NSE}(b) and (c) focus on the early and late times of an unstable breathing mode in this regime. At early times, the initial high-amplitude mode dominates the system and oscillates at approximately the frequency of the balanced case. This is highlighted in the power spectrum shown in the inset of \cref{fig:NSE}(b) with the frequency given by the vertical dashed line. As found in the self-evaporative regime, the energy of the droplet is dissipated through atom shedding which then leads to the evaporation of the initial mode. The oscillations at late times --- shown in \cref{fig:NSE}(c) --- are instead dominated by two other modes, with frequencies corresponding to the chemical potentials of each component given by the two outer vertical dashed lines. There are still residual oscillations from the decaying initial mode which hence explains the interference seen in the oscillations of \cref{fig:NSE}(c) and the central peak shown in the power spectrum. 

Thus, in both the self-evaporative and non-self-evaporative regimes, providing there is decay of the initial high-amplitude mode, there are two main regions of the dynamics: early times --- where the dynamics is dominated principally by a high-amplitude, intrinsic mode that is related to the balanced droplet; late times --- where there is a superposition of two modes corresponding to each chemical potential. All of these modes are decaying but over different timescales due to the rate of atom shedding. The initial mode decays relatively rapidly due to a high dissipation of energy from the particle emission. However, at late times the particle emission is considerably reduced, with slower emission of particles. This slower emission of particles, at energies of the two chemical potentials, is negligible relative to the droplet kinetic energy and hence the dynamics decay asymptotically \cite{ferioli2020}.

Finally, it should be noted that the breathing modes of imbalanced droplets have a larger range of instability than balanced droplets, which have an unstable breathing mode in the self-evaporative regime only. This is consistent with the discussions in both \cref{sec:gs} and the \cref{sec:var_deriv},of how the energy, and energy per particle, varies with imbalance. The key point from these discussions is that increasing imbalance corresponds to an increasing energy per particle and thus a less stably bound droplet. Hence, it would be expected that a less stably bound object would be more susceptible to particle shedding when perturbed, as is shown in the main results of this section.

\section{DISCUSSION AND CONCLUSIONS \label{sec:disc}}

By solving the spherically symmetric, coupled generalised GP equations, this work has systematically investigated the spherical ground states and breathing modes of imbalanced droplets across the size of the droplet and imbalance, in free space. Droplets can lower their energy by absorption of atoms, either by absorbing a symmetric number of atoms into each component, under the condition $N_2/N_1 = \sqrt{\gtt/\goo}$, or, as has been investigated here, if there exists an asymmetric quantity of atoms available, the droplet will absorb the available atoms forming a majority and minority component. This modifies the core structure of the droplet by splitting the two central component densities, yielding a bound, imbalanced droplet. If further majority component atoms become available then the imbalanced droplet saturates and thus any excess majority components will not bind to the droplet and reside outside of the droplet as an unbound gas. To investigate the density profiles, chemical potentials and breathing dynamics of these imbalanced droplets, without the external effects of the unbound gas, large computational boxes were used such that the background gas density is effectively zero.

One of the main experimental probes to justify the observation of a quantum droplet is measuring a constant width of the atomic cloud, after switching off all traps, i.e., the object is self-bound in free space \cite{semeghini2018,derrico2019,guo2021}. In this scenario any unbound gas would be lost, though the imbalanced droplet could still remain stable in free space as demonstrated in this work. The observation of the three breathing mode frequencies could be experimentally achieved in homonuclear mixtures, for which the number of atoms in each component can be tuned by radio-frequency pulses on the cloud \cite{cheiney2018}, and there have been recent experimental observations of imbalanced heteronuclear mixtures \cite{cavicchioli2022}. However these system will likewise be affected by three- and higher-body losses.

An immediate next question of this work is to explore how the application of trapping potentials effects imbalanced droplets. Isotropic harmonic traps could be applied to the spherically symmetric system considered here, to investigate the modification of both the ground states and breathing mode dynamics. The breathing mode dynamics of trapped, spherically symmetric droplets has been investigated in Ref.~\cite{hui2020} and so there are natural extensions of these results for imbalanced droplets.  Further ideas include investigating the spherically symmetric ground states and dynamics of a heteronuclear mixture, as the differing kinetic energy contributions of the two components may lead to novel dynamics. This is computationally challenging however, due to the form of the two-component LHY correction of a heteronuclear mixture \cite{petrov2015,ancilotto2018}. There are approximations for this term, though they are least accurate at the droplet centres \cite{minardi2019}. This presents an issue for accurate analysis of imbalanced ground state solutions and dynamics, especially for the superimposed, time-dependent collective excitations shown here. 

With the continuing development of quantum droplet experiments --- for example through the realisation of new mixtures \cite{wilson2020}--- it will become increasingly important to understand how these mixtures can minimise their energy, from self-evaporating excitations to modifying their internal structure.

The data presented in this paper are available \cite{taf_data}.

\begin{acknowledgments}
  The authors acknowledge support from the UK Engineering and Physical Sciences Research Council (Grant No. EP/T015241/1). T. A. F. also acknowledges support from the UK Engineering and Physical Sciences Research Council (Grant No. EP/T517914/1). The authors also thank Jakub Kopyci\'nski, Richard Tattersall and Nick Keepfer for useful discussions and patience.
\end{acknowledgments}

\appendix*
\section{Density difference of imbalanced, infinite droplets \label{sec:var_deriv}}
\begin{figure}[H]
  \centering
  \includegraphics[width=0.5\textwidth]{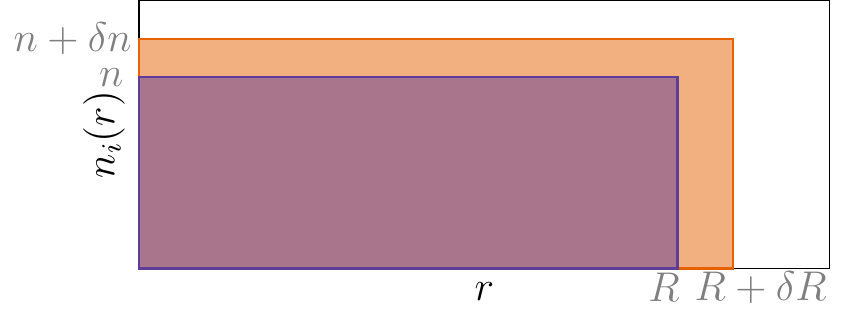}
\caption{Flat-top density profiles as an ansatz for each component of an imbalanced droplet. The orange component is the higher-density, majority component and the purple component is the lower-density, minority component.}
\label{fig:flat_top}
\end{figure}

The density profile of a droplet in the limit of large $\tilde{N}$ approaches a flat-top, step function, i.e., a constant density in the droplet core and a steep drop to zero density at the surface \cite{petrov2015}. This approximate form is a useful model for infinite sized droplets as it allows for kinetic energy contributions to be neglected. \cref{fig:flat_top} shows a schematic of the step function model of a large imbalanced droplet, used in this derivation. The minority component is modelled as having volume $V=\frac{4}{3}\pi R^3$, whilst the majority component has volume $V+\delta V = \frac{4}{3}\pi(R+\delta R)^3$. The step functions hence correspond to the central densities $n$ and $n+\delta n$ for the minority and majority components, respectively, with the two components normalised to $N$ and $N+\delta N$ in which $\delta N\geq0$, imposing the population imbalance.
The dimensionless form of the energy functional given in \cref{eq:e_func}, with $\beta \equiv \sqrt{\att/\aoo} = 1$, can be written in terms of these constant imbalanced densities giving,
\begin{equation}
  \begin{split}
    E = \int&\Bigg[\frac{(n+\delta n)^2}{2}+ \frac{n^2}{2} + \eta (n+\delta n)n \\
      &+ \frac{2}{5}\alpha\left(2n + \delta n\right)^{5/2}\Bigg]\dd^3\mathbf{r}.
  \end{split}
  \label{eq:dens_energ}
\end{equation}
The schematic in \cref{fig:flat_top} states nothing about the sign of $\delta R$, i.e., the volume of the majority component, $V + \delta V$, can be larger or smaller than the minority component volume, $V$. However, it can be shown that the only physical system is that where $\delta V = 0$, which is demonstrated in the following calculation.

Beginning with the $\delta V > 0$ case, \cref{eq:dens_energ} can be written in terms of the component volumes and population numbers as,
\begin{equation*}
  \begin{split}
    E & = \frac{(N+\delta N)^2}{2(V+\delta V)} + \frac{N^2}{2V} + \eta\frac{N(N+\delta N)}{V+\delta V} \\
      & + \frac{2}{5}\alpha\left\{V\left[\frac{N+\delta N}{V+\delta V} + \frac{N}{V}\right]^{5/2} + \delta V\left[\frac{N+\delta N}{V +\delta V}\right]^{5/2}\right\}.
\end{split}
\end{equation*}
To analyse how the energy of this model varies with $\delta V$, the energy above can be minimised with respect to the difference in component volumes, giving,
\begin{equation*}
  \begin{split}
    \pdv{E}{(\delta V)} & = -\frac{(N+\delta N)^2}{2(V+\delta V)^2} - \eta\frac{N(N+\delta N)}{(V+\delta V)^2} \\
                        & + \frac{2}{5}\alpha\Bigg\{-\frac{5}{2} \frac{V(N+\delta N)}{(V+\delta V)^2}\left[\frac{N+\delta N}{V+\delta V} + \frac{N}{V}\right]^{3/2}  \\
                        &+ \left[\frac{N+\delta N}{V + \delta V}\right]^{5/2} - \frac{5}{2} \delta V\frac{(N+\delta N)^{5/2}}{(V +\delta V)^{7/2}}\Bigg\}.
  \end{split}
\end{equation*}
Taking the limit of $\delta V \rightarrow 0^{+}$ reduces to,
\begin{equation*}
  \begin{split}
    \pdv{E}{(\delta V)}\eval_{\delta V\rightarrow0^{+}} & = -\frac{(N+\delta N)^2}{2V^2} - \eta\frac{N(N+\delta N)}{V^2} \\
                        & + \alpha\Bigg\{-\frac{(N+\delta N)}{V}\left[\frac{2N+\delta N}{V}\right]^{3/2} \\
                        &+ \frac{2}{5}\left[\frac{N+\delta N}{V}\right]^{5/2}\Bigg\},
\end{split}
\end{equation*}
which can be written in the form,
\begin{equation*}
  \pdv{E}{(\delta V)}\eval_{\delta V\rightarrow0^{-}} = -\frac{N+\delta N}{2V^2}\left[(1+2\eta)N + \delta N\right] + \mathcal{O}(\alpha).
\end{equation*}
Recall from \cref{sec:mod} that $\delta a = \aot + \sqrt{\aoo\att}$, and that defining $\delta a <0$ implies dominantly attractive interactions, which can be written as $\eta<-1$ in the dimensionless parameters used here. Thus, with $\delta N$ small, the mean-field terms becomes dominantly positive whilst the LHY contributions are small due to $\alpha \ll 1$. Hence, the limit of $\delta V \rightarrow 0^{+}$ yields, $\pdv*{E}{(\delta V)} \geq 0$. 

The next case to check is $\delta V<0$, for which \cref{eq:dens_energ} takes the form,
\begin{equation*}
  \begin{split}
    E & = \frac{(N+\delta N)^2}{2(V+\delta V)} + \frac{N^2}{2V} + \eta\frac{N(N+\delta N)}{V} \\
      & + \frac{2}{5}\alpha\left\{(V+\delta V)\left[\frac{N+\delta N}{V+\delta V} + \frac{N}{V}\right]^{5/2} - \delta V\left[\frac{N}{V}\right]^{5/2}\right\}.
\end{split}
\end{equation*}
This expression is again minimised with respect to the difference in component volumes,
\begin{equation*}
  \begin{split}
    \pdv{E}{(\delta V)} & = -\frac{(N+\delta N)^2}{2(V+\delta V)^2} + \frac{2}{5}\alpha\Bigg\{\left[\frac{N+\delta N}{V+\delta V} + \frac{N}{V}\right]^{5/2} \\
                        & -\frac{5}{2}\frac{(N+\delta N)}{V+\delta V}\left[\frac{N+\delta N}{V + \delta V} + \frac{N}{V}\right]^{3/2} - \left(\frac{N}{V}\right)^{5/2}\Bigg\}.
  \end{split}
\end{equation*}
Taking the limit of $\delta V \rightarrow 0^{-}$ yields,
\begin{equation*}
  \begin{split}
    \pdv{E}{(\delta V)}\eval_{\delta V\rightarrow0^{-}} & = -\frac{(N+\delta N)^2}{2V^2} + \alpha\Bigg\{\frac{2}{5}\left[\frac{2N+\delta N}{V}\right]^{5/2} \\
                        & - \frac{N+\delta N}{V}\left[\frac{2N+\delta N}{V}\right]^{3/2} - \frac{2}{5}\left(\frac{N}{V}\right)^{5/2}\Bigg\},
\end{split}
\end{equation*}
which can be written in the form,
\begin{equation*}
  \pdv{E}{(\delta V)}\eval_{\delta V\rightarrow0^{-}} = -\frac{(N+\delta N)^2}{2V} + \mathcal{O}(\alpha).
\end{equation*}
Hence, the limit of $\delta V \rightarrow 0^{-}$ results in, $\pdv*{E}{(\delta V)} \leq 0$. Thus, the conditions
\begin{equation*}
  \pdv{E}{(\delta V)}\eval_{\delta V\rightarrow0^{+}} \geq 0, \, \mathrm{and}, \, \pdv{E}{(\delta V)}\eval_{\delta V\rightarrow0^{-}} \leq 0,
\end{equation*}
imply that $\delta V=0$ is either a smooth minimum or a cusp. This is physically realistic in the limit of large droplets, as this implies that the radii of the two components are equal, i.e., no single-component atoms reside outside of the droplet core. However, in the finite droplet limit this is not the case as kinetic energy contributions cannot be neglected, as is highlighted in the ground state profiles in \cref{fig:gs_profs} (b) and (c). The imbalanced droplet hence appears to slowly approach the flat-top density profile, relative to the rate at which the large, balanced droplet approaches the flat-top density limit.

Taking the case of $\delta V = 0$, the energy can be written as,
\begin{equation*}
  \begin{split}
    \frac{E}{2N} &= \frac{(N+\delta N)^2}{4NV} + \frac{N}{4V} + \eta\frac{N+\delta N}{2V} \\
  &+\frac{2}{5}\alpha\left[\frac{V}{2N}\left(\frac{2N+\delta N}{V}\right)^{5/2}\right].
  \end{split}
\end{equation*}
A factor of $\left(1+\delta N/2N\right)$ can be factored out of this expression, allowing for the above equation to be written in powers of $(N/V)$,
\begin{equation*}
  \frac{E}{2N} = \left(1 + \frac{\delta N}{2N}\right)\left[\mathcal{A}\left(\frac{N}{V}\right) + \mathcal{B}\left(\frac{N}{V}\right)^{3/2}\right],
\end{equation*}
where
\begin{equation*}
  \mathcal{A} = \left(1+\frac{\delta N}{2N}\right) + (\eta-1) + \frac{1-\eta}{2}\left(\frac{1}{1+\frac{\delta N}{2N}}\right),
\end{equation*}
and 
\begin{equation*}
  \mathcal{B} = \frac{4\sqrt{2}\alpha}{5}\left(1 + \frac{\delta N}{2N}\right)^{3/2}.
\end{equation*}
\cref{sec:mod} discusses the key property that droplets exist in equilibrium with the vacuum, i.e., they have zero pressure, $\pdv*{E}{V} = 0$. With $N$ constant, this expression for the zero pressure droplet can be rewritten as $\pdv*{(E/2N)}{(N/V)}=0$, yielding,
\begin{equation*}
  \left(\frac{N}{V}\right)^{1/2} = -\frac{2\mathcal{A}}{3\mathcal{B}} \implies n=\left(\frac{N}{V}\right) = \frac{4\mathcal{A}^2}{9\mathcal{B}^2},
\end{equation*}
giving an expression for the equilibrium density. This results in the equilibrium energy, given by,
\begin{equation}
  E_{\mathrm{eqbm}} = \left(2N+\delta N\right)\left[\frac{4\mathcal{A}^3}{27\mathcal{B}^2}\right],
  \label{eq:var_energy}
\end{equation}
or equivalently as the equilibrium energy per particle,
\begin{equation}
\left(\frac{E}{2N+\delta N}\right)_{\mathrm{eqbm}} = \frac{4\mathcal{A}^3}{27\mathcal{B}^2}.
  \label{eq:var_energy_perp}
\end{equation}
For fixed $\alpha$, $\eta$ and $N$, Equations \ref{eq:var_energy} and \ref{eq:var_energy_perp} are plotted as functions of $\delta N$ in \cref{fig:var_fig}. The main plot in \cref{fig:var_fig} shows \cref{eq:var_energy} varying with $\delta N$. By differentiating \cref{eq:var_energy}, the minimum is located at
\begin{equation*}
  \frac{\delta N}{N} = 2 (\eta - 2) + [(\eta - 1) (4 \eta - 14)]^{1/2}.
\end{equation*}
For the parameters of \cref{fig:var_fig} the minimum is at $\delta N \approx 703$, which using the expression,
\begin{equation*}
  \delta n = \frac{\delta N}{N}\left(\frac{4\mathcal{A}^2}{9\mathcal{B}^2}\right),
\end{equation*}
corresponds to $\delta n\approx0.458$, i.e., the orange, dashed, horizontal line in the inset of \cref{fig:gs_profs}(d). 


\bibliography{references.bib}

\end{document}